\documentclass[%
reprint, showpacs,
 amsmath,amssymb,
 aps,
prl,
]{revtex4-2}

\usepackage{graphicx}
\usepackage{dcolumn}
\usepackage{bm}
\usepackage{subfigure}
\usepackage{color}
\usepackage{lipsum}
\usepackage{verbatim}

\newcommand{\ve}[1]{{\bf{#1}}}

\newcommand{\matr}[1]{\underline{\underline{#1}}}
\newcommand{\deriv}[2]{\frac{\partial {#1}}{\partial {#2}}}
\newcommand{\vi}[1]{\mathbf{#1}}

\newcommand{\sinf}  {\sum_{n=-\infty}^{\infty}}

\def\be#1\ee{\begin{equation}#1\end{equation}}
\def\bea#1\eea{\begin{align}#1\end{align}}
\def\bse#1\ese{\begin{subequations}#1\end{subequations}}

\def\1#1{{\hat{{\boldsymbol{#1}}}}}                                 		
\def\2#1{\hat{#1}}                                              		   		
\def\3#1{{\mathbf{#1}}}                                             	   		
\def\4#1{{\boldsymbol{#1}}}                                            		
\def\5#1{{\mathcal#1}}                                                            		
\def\6#1{\bar{#1}}                                                           		
\def\7#1{{{#1}}}                                    		
\def\8#1{\widetilde{#1}}
\def\9#1{\check{#1}}
\def\+#1{{\overset{{\scriptscriptstyle +}}{#1}{}}}                  		
\def\b+#1{{\overset{{\scriptscriptstyle +}}{\mathbf{#1}}{}}}        	
\def\g+#1{{\overset{{\scriptscriptstyle +}}{\boldsymbol{#1}}{}}}    

\definecolor{dark-green}{rgb}{0.278,0.7,0.4}                    

\definecolor{my-brown}{rgb}{0.69,0.247,0.13}                    

\definecolor{my-purple}{rgb}{0.47,0.12,0.46}                    

\definecolor{my-greenblue}{rgb}{0.129,0.313,0.419}              

\definecolor{my-orange}{rgb}{1,0.5,0.25}                        

\definecolor{my-red}{rgb}{0.745,0,0.2117}                       

\definecolor{my-gray}{rgb}{0.5,0.5,0.5}                         

\definecolor{my-dark-blue}{rgb}{0.1,0.1,0.7}                    

\definecolor{my-indigo}{rgb}{0.29,0.0,0.51}                    

\begin{document}

\preprint{APS/123-QED}




\title{One-way Acoustic Guiding under Transverse Fluid Flow}

\author{Ohad Silbiger and Yakir Hadad}
\email{hadady@eng.tau.ac.il}
\affiliation{School of Electrical Engineering, Tel-Aviv University, Ramat-Aviv, Tel-Aviv, Israel, 69978}

\date{\today}

\begin{abstract}
In a moving acoustic medium, sound waves travel differently with and against the fluid flow. This well-established acoustic  effect is backed by the intuition that the fluid velocity bias imparts momentum on the propagating acoustic waves, thus violating reciprocity.  Based on this conception, fluid flow that is transverse to the wave direction of propagation will not break reciprocity.
%
%
In this letter we contrast this common wisdom and theoretically show that the interplay between transverse mean flow and transverse structural gliding-asymmetry can yield strong nonreciprocity and even, surprisingly, one-way  waveguiding which is a rare in acoustics.
\end{abstract}

\pacs{Valid PACS appear here}
\maketitle



\begin{figure}[h]
    \includegraphics{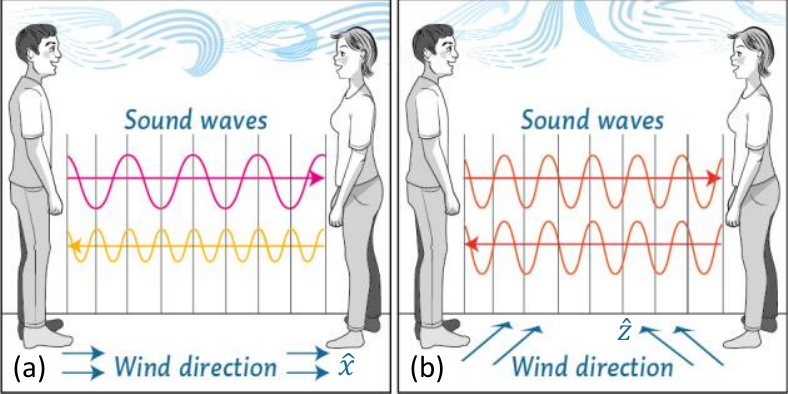}
    \centering
    \caption{(a) When the medium between Alfred and Beth flows parallel to the plane waves that propagate between them, different behavior is expected when the propagation is from left to right or vice versa. Since  only the medium is flowing, while Alfred and Beth are stationary, the wave \emph{frequency} as observed by them does not experience a Doppler-shift. However, due to the medium flow, the \emph{wavelengths} of the two counter-propagating waves are different \cite{Morse}. (b) In contrast, if the medium between Alfred and Beth flows transversely to the plane waves that propagate between them, these plane waves are expected to propagate reciprocally \cite{Comm1}.}
    \label{fig:ab}
\end{figure}

\textit{Introduction}.\textemdash
Acoustic nonreciprocity has gained a lot of attention in recent years due to its numerous potential applications \cite{FleuryReview, NassarReview}. It  can be achieved using nonlinearities  \cite{Liang2009, Boechler2011, Cui2018}, and by using active elements \cite{Popa2014, Guo2020, Zhai2019}, but nevertheless,
nonreciprocal acoustic and phononic propagation is mostly known to occur in moving media \cite{Godin1997, Wiederhold2019, Morse, Fleury2014, Auregan2017, Auregan2015}, or some other form of internal motion such as rotation \cite{Wang2015}. To illustrate that, assume that two friends, Alfred and Beth, shown in Fig.~\ref{fig:ab}(a), are located at points A and B on the $x$ axis,  the distance between them $L$, is acoustically large, and under this high frequency limit they communicate via \emph{local} plane waves that travel along the $x$ axis in a fluid with uniform mean flow velocity $U_0\hat{x}$, and acoustic wave speed $c$.   Using simple kinematic arguments, the time it takes for a signal from Alfred to reach Beth is different then the time it takes for a signal to propagate in the reciprocal direction: $ \tau_d^{A\leftrightarrow B} = L/(c \pm U_0) $. Clearly, in light of this result, nonreciprocity that is caused due to a collinear fluid flow is rather weak at low Mach numbers, $q_0=U_0/c\ll1$. Nevertheless, in the presence of sharp resonances the effect may be enhanced giving rise even to  isolation. This has been demonstrated, for example, using resonant cavities \cite{Fleury2014, Ding2019, Yang2015, Fleury2015},  waveguides connected in a sensitive Mach-Zehnder interferometer setup \cite{Jankovic2020} and near zero index metamaterial waveguides \cite{Quan2019}. Moreover,  faster \emph{synthetic} motion and thereby stronger nonreciprocity, may be emulated by space-time modulated acoustic metamaterials \cite{Wang2018, Croenne2019, Huang2018, Nassar2017, Swinteck2015, Fleury2016}.
In contrast, if, as illustrated in Fig.~\ref{fig:ab}(b), the fluid flows \emph{transverse} to the direction of propagation, so the flow velocity is $U_0\hat{z}$, the time delay for a \emph{local plane wave} to propagate between Alfred and Beth is $\tau_d^{A\leftrightarrow B} = L/c$, independent of the direction of propagation, as in a stationary fluid \cite{Comm1}. The exact results for time of flight in both cases (longitudinal vs transverse flow) are highly intuitive from the kinematic point of view which implies, allegedly, that in the presence of a uniform mean flow, the communication between Alfred and Beth will be nonreciprocal only if the wave that propagates between them has some wavevector component that is parallel to the fluid stream.
%

In contrast with this common wisdom, in this letter we show that under certain conditions it is possible to achieve strong nonreciprocity also for waves that are propagating along a direction that is \emph{transverse} to the uniform flow velocity of the ambient medium. As we show below, this is achieved by the interplay between the transverse flow and structural the transverse gliding-asymmetry of acoustic scatterers. Interestingly, this effect can be explained only by considering the \emph{complete field} interaction of the scatterers and its effect on the travelling mode.
Remarkably, by this way we demonstrate below strong nonreciprocity and even  \emph{one-way guiding} that is due to \emph{transverse flow} biasing.  In a striking contrast with the common perception that is illustrated in Fig.~\ref{fig:ab}(b).

\begin{figure}[h]
    \includegraphics{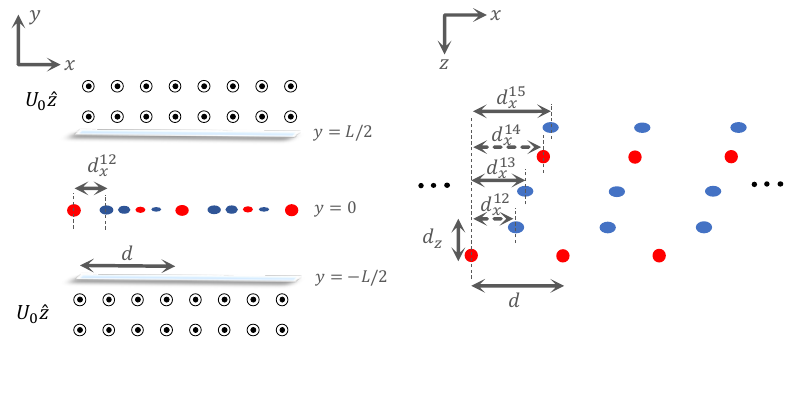}
    \centering
    \caption{A periodic waveguide with transverse asymmetry (left - side view, right - top view). As an example, the waveguide consists of five chains of acoustic scatterers, and is located inside a \emph{stationary} medium, sandwiched between two half-spaces with mean transverse flow with velocity $U_0\hat{z}$. $d$ is the inter-scatterer spacing along each one of the chains, and $d_z$ is the spacing between the chains. The lattice gliding-asymmetry is determined by the chain gliding parameters - $d_x^{1j}$, $j=2,3,4,5$. The waveguide is infinite in the $x$ direction, only three unit cells are shown.
    }
    \label{fig:struct}
\end{figure}

\textit{Description of the model}.\textemdash
We consider a coupled resonator waveguide that comprises of $N$ periodic linear chains of scatterers. The waveguide is located on the $y=0$ plane, parallel to the $x$-axis, and consists of a periodic arrangement of scatterers that are equally spaced with acoustically-small inter-scatterer spacing $d<\lambda/2$, where $\lambda$ is the acoustic wavelength. See illustration in Fig.~\ref{fig:struct} with $N=5$.
%
%
%
The $n$'th scatterer of chain number $i$ ($i=1..N$) is located at $\mathbf{r}_{n}^{i}=(d_x^i+nd,0,d_z^i)$. Here, $d_z^i=(i-1)d_z$ denotes the $z$ coordinate of the $i$'th chain, where $d_z^1=0$ so that the first ($i=1$) chain  is located right on the $x$-axis, and the spacing between the chains is $d_z$. And, $d_x^i$ is the \emph{gliding} distance of the $i$'th chain with respect to the origin. For the first chain,  $d_x^1=0$. Thus, the relative gliding between the chains reads $d_x^{ij} = d_x^j - d_x^i$.
The entire structure is located inside a duct with a \emph{quiescent} (stationary) medium that is sandwiched between two half-spaces with an uniform flow with velocity $U_0\hat{z}=q_0c_0\hat{z}$, \emph{transverse} to the coupled resonator waveguide axis. We assume that the density, $\rho_0$, and the wave velocity, $c_0$ are uniform in the entire space. Thus, the duct is created merely due to the inhomogeneous medium flow velocity.
This configuration may be achieved in practice using a thin, acoustically transparent, membranes placed at the interfaces, $y=-L/2$ and $y=L/2$, as illustrated in the left panel of Fig.~\ref{fig:struct}. We stress that by using this duct configuration we completely \emph{avoid the possible issue of turbulent flow} at high Raynolds numbers which could yield a porus-media like behaviour that may be expected by the presence of the acoustically dense scatterers, had the flow was applied in the \emph{entire} space. This implies that in the suggested configuration, the medium flow lines are necessarily not affected by the  lattice of scatterers, and hence they are fully transverse to the waveguide axis.
%
%
The boundary condition in the scenario suggested in Fig.~\ref{fig:struct}, at the interfaces $y=\pm L/2$, has been extensively studied, for example in \cite{Miles1957}, and experimentally explored in \cite{AMIET1975}.

\textit{Modelling the acoustic scatterers}.\textemdash
We assume that the acoustic scatterers may exhibit either dominant monopole or dipole response.
These scatterers may be passive, for example Helmholtz resonators \cite{Morse} and variations over them \cite{Quan2018}, where the structural properties of the scatterer determine its response, or alternatively, convenient tunable acoustic meta-atoms may be utilized \cite{Popa2013}. The latter are essentially active scatterers that  consist of a transducer that senses the pressure wave incident on the meta-atom and an electronic feedback circuit that manipulates the electric signal produced and drives a second transducer that creates the desired acoustic response which may be of a monopole scatterer, a dipole scatterer, and in principle of any multipole.
Thus, for the sake of generality, we assume that
the scatterers are characterized by their acoustic susceptibilities, $\alpha^{mm}$ and $\alpha^{dd}_{\xi}$, which link the \emph{local field}, i.e., the field in the scatterer's location but at the absence of the scatterer itself, to the resulting scattering response. Specifically, the monopole is characterized by a volume $V$ (units: $m^3$) and is induced by a local pressure $P$, while the dipole is characterized by a dipole moment $D_{\xi}$ (units: $m^4$) and is induced by the space derivative of the local pressure $\partial P/\partial \xi$, in the direction $\xi=x,y,z$. Thus,
\begin{equation}\label{eq:suscept}
    V = \alpha^{mm} P \qquad , \qquad D_{\xi} = \alpha^{dd}_{\xi} \deriv{P}{\xi}.
\end{equation}
%
We use a Lorentzian model for the susceptibilities, i.e., $1/\alpha = A(\omega^2-\omega_r^2)-j\omega\gamma$ where $\omega_r$ denotes the resonance frequency, $A$ takes the role of the oscillator strength, and $\gamma$ encapsulates the resonator damping. Here $\alpha$ stands for either $\alpha^{mm}$ or $\alpha_{\xi}^{dd}$ for monopole and dipole scatterers, respectively. Specifically, the radiation loss for a monopole scatterer is given by $\gamma=\gamma^{mm}=\omega^2\rho_0/4\pi c$, while for a dipole scatterer it reads  $\gamma=\gamma^{dd}_{\xi}=\omega^4\rho_0/12\pi c^3$ (see \cite{Mazor2020} and Sec. I in \cite{SM}). Additional loss mechanisms may be added to $\gamma$. Specific values for the resonators' resonance frequency and strength are given in the numerical examples below.
Once the \emph{induced source} on the scatterer is known, the scattered field generated by the scatterer is given via the corresponding Green's function that takes into account the wave interaction at the boundaries with the uniform flow domains.
As an example, for a monopole scatterer that is located at $\ve{r}'=(0,0,0)$, the scattered acoustic pressure on the $y=0$ plane is given by $P_{scat}(\ve{r})=G^{mm}(\ve{r})V$, where \cite{SM} (Sec. II)
\begin{equation} \label{eq:Gmm_spectrum}
    G^{mm}(\ve{r})\! =\! \iint_{-\infty}^{\infty}\!\! g^{mm}(k_x, k_z) e^{-j(k_x x + k_z z)} dk_x \, dk_z.
\end{equation}
The spectral green's function in Eq.~(\ref{eq:Gmm_spectrum}) reads,
\begin{equation} \label{eq:gmm_spectral2}
    g^{mm}(k_x, k_z) = \frac{j\omega^2 \rho_0}{8\pi ^2 k_y}
                           \frac{ 1 + R e^{-j k_y L} }{ 1 - R e^{-j k_y L} }
\end{equation}
with
\begin{equation}
    R = \frac{k_2^2k_{y1}-k_1^2k_{y2}}{k_2^2k_{y1}+k_1^2k_{y2}}
\end{equation}
and where $k_{y1,2}=\sqrt{k_{1,2}^2-k_x^2-k_z^2}$, with $k_1=\omega/c_0$ and $k_2=(\omega-U_0 k_z)/c_0$,  in the stationary medium and in the uniform flow domains, respectively.
Additional Green's functions, for different fields, and due to additional sources, are listed in Ref.~\cite{SM}.

\textit{Guided modes}.\textemdash
Given a specified spatial configuration of the waveguide, such as in Fig.~\ref{fig:struct}, our interest is to find the guided modes that may propagate through it. To that end, we consider an infinite lattice.
%
Due to its structural periodicity,  the \emph{induced sources on the scatterers} have to respect a Bloch-form, i.e., exhibiting a \emph{collective} response with the following wave behavior: $X^i_n = X^i_0  e^{-j\beta nd}$, where $\beta$ is the unknown propagation constant of the eigenmode.
Henceforth we use $X_n^i$ to denote the induced source moment on the $n$'th scatterer in chain number $i$, which may be a monopole ($V$) or a dipole ($D_x$ or $D_z$).
%
%
%

Given two scatterers, with indexes $n$ and $m$ on chains number $i$ and $j$, i.e., $X_n^i$ and $X_m^j$, respectively, that are located at $\vi{r}_n^i=(d_x^i+nd,0,d_z^i)$ and $\vi{r}_m^j=(d_x^j+md,0,d_z^j)$ on the $y=0$ plane (See Fig.~\ref{fig:struct}). We denote by $G^{ij}(\vi{r}_n^i,\vi{r}_m^j)$ the Green's function that relates the induced source $X_n^i$ to the \emph{field that excites} $X_m^j$. For example, if the scatterers are both monopoles, i.e.,  $X_n^i=V_n^i$, $X_m^j=V_m^j$, then $G^{ij}$ is defined as the Green's function that connects a monopole source to the excited pressure wave, that is, $P(\vi{r}_n^i)=G^{ij}(\vi{r}_m^j,\vi{r}_n^i)V_n^i$.
Note that since all the scatterers are located on $y=0$, and since the duct structure (in the absence of the scatterers) is shift invariant on the $xz$ plane, then $G^{ij}(\vi{r}_m^j,\vi{r}_n^i)=G^{ij}(\vi{r}_m^j-\vi{r}_n^i)$.

Next, we close the loop, and write the equations that govern the excitation of the scatterers on chain number $j$. In light of the structural periodicity, it is enough to write the equation for $X_0^j$. Using its susceptibility, and using the local field at its location, $E^L(\ve{r}_0^j)$,
\begin{equation}\label{eq:local field}
X_0^j=\alpha^jE^L(\ve{r}_0^j)=\alpha^j \sum_{i=1}^N S^{ij}X_0^i.
\end{equation}
Here, $S^{ij}X_0^i$ denotes the field at $\vi{r}_0^j$ due to all the scatterers at chain number $i$. $S^{ij}$ is expressed in terms of $G^{ij}$,
\begin{equation}\label{eq:summation}
    S^{ij}(\beta) = \sinf e^{-j\beta nd} \, G^{ij}(\vi{r}_0^j-\vi{r}_n^i).
\end{equation}
where $\vi{r}_m^j-\vi{r}_n^i=(d_x^{ij} - nd, 0, d_z^{ij})$. Note that  $G^{ij}(\vi{r}_0^j-\vi{r}_n^i)$ consists of \emph{two contributions}, a \emph{primary} wave that would be excited in an infinite homogenous medium with no flow, and a \emph{secondary} wave that is reflected by the boundaries with the moving medium.
Importantly, for $i=j$, and when $n=0$ in Eq.~(\ref{eq:summation}), $G^{ii}(\vi{r}_0^i-\vi{r}_0^i)$ consists of only the secondary wave field. This is because the local field $E^L(\ve{r}_0^j)$ is by definition the field at the particle location but in the absence of the particle \emph{itself}. Reradiation effect of the scatterer on itself though the boundaries are essential. The numerical evaluation of Eq.~(\ref{eq:summation}) is discussed in Ref.~\cite{SM} (Sec. III).
%
%
By rewriting Eq.~(\ref{eq:local field}) for all $j=1..N$, a self-consistent linear system that encapsulates the  modal dynamics of the waveguide with $N$ parallel chains is obtained,
\begin{equation}\label{eq:latticedynamics}
\vi{X}_0=\matr{\alpha}(\omega)\matr{S}(\omega,\beta)\vi{X}_0
\end{equation}
where $\vi{X}_0=[X_0^1,..,X_0^N]^T$, $\matr{\alpha}=\mbox{diag}[\alpha^1,..,\alpha^N]$, and $\matr{S}=\{S^{ij}\}$ with $i=1..N,j=1..N$.
%
%
The waveguide modes are the nontrivial solutions of Eq.~(\ref{eq:latticedynamics}) and as such  satisfy the dispersion equation
\begin{equation}\label{eq:dispersion}
\mbox{det}[\matr{\alpha}(\omega)^{-1} - \matr{S}(\omega, \beta)]=0
\end{equation}
that its solutions are $\beta(\omega)$.
Real $\beta$ solutions will exist only if $d<\lambda/2$.
Therefore implying sub-wavelength mode width as expected by a sub-diffraction waveguide (see \cite{Quinten1998, Alu2006, Hadad2010} for akin waveguides in optics).
%

\textit{Nonreciprocity and one way guiding}.\textemdash
In the absence of flow, $q_0=0$, reciprocity is expected. Mathematically, this is evident by the symmetry (or anti-symmetry) of all Green's functions $G^{ij}$ with respect to $x$ and $z$ which implies $\matr{S}(-\beta) = \matr{S}^T(\beta)$. Therefore, in light of the form of  Eq.~(\ref{eq:dispersion}), for any given frequency $\omega$, if $\beta$ solves the dispersion equation,  so does $-\beta$.
The same argument is true even in the presence of transverse flow, $q_0\neq0$, if the waveguide contains \emph{only zero or half-step gliding} for any number of chains, i.e., if $\forall i \neq j \quad d_x^{ij} = 0$ or $d_x^{ij} = d/2$. In this case the waveguide's transverse spatial symmetry enforces reciprocal wave guiding. Detailed derivation of the symmetry conditions is in \cite{SM} (Sec. IV). %

Instead, as we show below, it requires to have \emph{simultaneously transverse flow and spatial asymmetry} in the form of gliding in order to breach reciprocity and get different propagation characteristics for counter propagating  waves along the waveguide axis.
%
%
%
%
%
\begin{figure}[h]
    \includegraphics{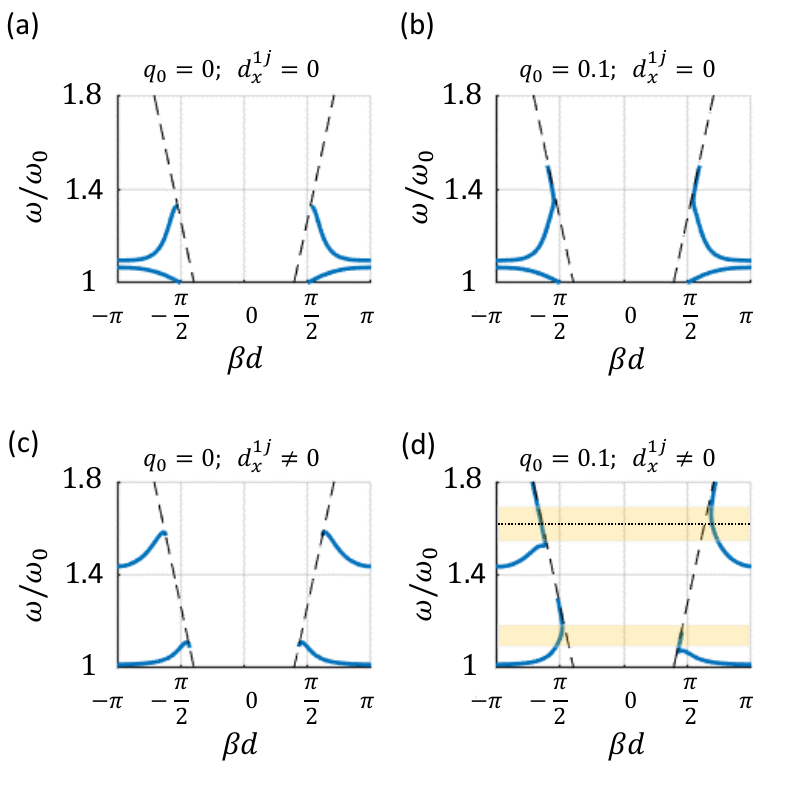}
    \centering
    \caption{Dispersion plots for a waveguide comprised of 5 scatterer chains - two chains of monopoles and three chains of longitudinal dipoles. (a) With no flow above and below the duct and no chain gliding the propagating mode is symmetric, both solutions have equal phase and group velocities, but in opposite directions. (b) Adding transverse flow while maintaining zero gliding changes the dispersion plots very slightly, without affecting the reciprocal nature of the waveguide. (c) Introducing chain gliding in a quiescent medium changes the dispersion curves, but does not break reciprocity. (d) Combination of chain gliding-asymmetry and transverse flow generates substantial nonreciprocity in the waveguide.}
    \label{fig:dispersion}
\end{figure}
%
%
The aforementioned discussion is summarized by the dispersion plots that are given in Fig.~\ref{fig:dispersion}. In this case the structure consists of total of five coupled linear arrays, two with monopole scatterers and three with longitudinal dipole scatterers. The linear arrays are shifted relative to each another to produce a gliding asymmetry. The medium is uniformly characterized as air with $c=343$m/s and $\rho_0=1.2\mbox{kg/m}^3$. The monopoles resonate at a frequency of $f^{mm}_r=f_0=1125$Hz, while the dipoles exhibit no resonance around this frequency, but around $f^{dd}_{z}=3377$Hz. The resonator strength for the monopole $A^{mm}=10\mbox{kg/m}^4$, and for the dipoles  $A^{dd}_x=10^{-3}\mbox{kg/m}^6$.  See \cite{SM} (Sec. I) for susceptibility plots.
The thickness of the layer between the flowing media was set to quarter of a wavelength
at frequency $f_0$, that is $L=7.5$cm in Fig.~\ref{fig:struct}. The lattice periodicity is taken to be $d=6$cm, the transverse inter-chain spacing is  $d_z=2$cm, and the gliding values are: $d_x^{12}=0.66d$, $d_x^{13}=0.5d$, $d_x^{14}=0.33d$, and $d_x^{15}=0.33d$.
By solving  Eq.~(\ref{eq:dispersion}) we find the dispersion relation $\omega(\beta)$ that is plotted in Fig.~\ref{fig:dispersion}. In the absence of medium flow ($q_0 = 0$) \emph{or} when the structure is transversely symmetric (for all $i,j$ $d_x^{ij} = 0$ or $d/2$), the dispersion relation is symmetric (in $\beta$), as shown in Fig. \ref{fig:dispersion}(a)-(c), and the waveguide is reciprocal. It is only when transverse flow is introduced simultaneously with a transversely gliding-asymmetric structure, then reciprocity in the longitudinal direction is broken and the dispersion relation becomes asymmetric as shown in Fig.~\ref{fig:dispersion}(d) with Mach number $q_0=0.1$.
In this case, remarkably, we observe certain frequency bands (See the colored regions in Fig.~\ref{fig:dispersion}(d)) at which a real $\beta$ solution exits only for modes that are propagating to one side. At these bands the transversely biased waveguide acts as a longitudinal one-way waveguide.
We note that these one-way guiding bands in the dispersion cannot be found by using a mere kinematic calculation. This is demonstrated in detail in \cite{SM}(Sec. V) using a coupled mode model with non-reciprocal coupling coefficients due to the transverse flow. Instead, a detailed calculation that takes carefully into account the complete acoustic field form is essential.

\begin{figure}[h]
    \includegraphics{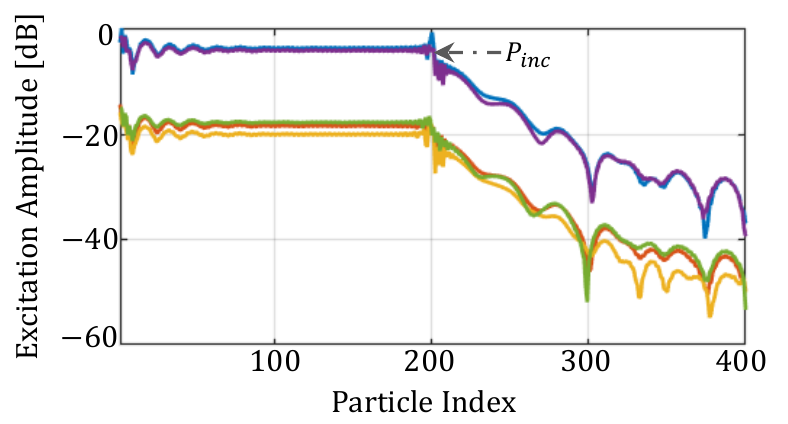}
    \centering
    \caption{Amplitudes of monopole/dipole strength along a chain with 400 scatterers, due to a local external pressure field $P_{inc}$ applied on a single monopole scatterer in the middle of chain $i=1$. Color coded: blue - $V_n^1$, orange - $D_{z_n}^2$, yellow - $D_{z_n}^3$, magenta - $V_n^4$, green - $D_{z_n}^5$. The lattice parameters used are the same as in Fig.~\ref{fig:dispersion}(d), with Mach number $q_0=0.1$ and at frequency $\omega=1.6\omega_0$. Only propagation to the left is allowed consistent with the negative group velocity obtained from the dispersion diagram in Fig.~\ref{fig:dispersion}(d) at this frequency.}
    \label{fig:finite}
\end{figure}

To further demonstrate the phenomenon and validate our dispersion results, a finite waveguide with 400 unit-cells is excited by applying a \emph{localized} external pressure field that oscillates at $\omega=1.6\omega_0$ on a monopole scatterer in the middle of chain $i=1$.  At this excitation frequency, with flow velocity with $q_0=0.1$, the dispersion diagram in Fig.~\ref{eq:dispersion}(d) predict a single real solution with negative group velocity ($d\omega/d\beta$). Thus, we expect in the excitation problem to see propagation to the left, and an evanescent wave to the right.
The numerical excitation results of the finite lattice are shown in Fig.~\ref{fig:finite}, while the numerical details are omitted here for brevity and are provided in \cite{SM} (Sec. VI).
%
%
These results nicely demonstrate how under transverse mean flow, longitudinal one-way leftward propagation of the acoustic wave is obtained, while rightward propagation is practically prohibited. Instead, rapid exponential decay takes place in the forbidden direction, indicating on a complex leaky mode that is excited in this direction.  A detailed analysis of this excitation problem should follow Green's function development for the waveguide under  study here. Related work may be found in \cite{Hadad2011}, and in \cite{Hadad2013}. The latter discusses  the Green's function of a one-way plasmonic sub-diffractive waveguide under magnetic biasing.
Note that additional examples using a waveguide lattice that comprises of monopoles only, and including additional loss terms are given in \cite{SM}(SEc. VII).

\textit{Conclusions}.\textemdash
We have theoretically shown that an acoustic waveguide that is placed  in the \emph{vicinity} of a flow with uniform velocity \emph{transverse} with respect to the waveguide axis can exhibit substantial nonreciprocity and even one-way guiding. This counter-intuitive phenomenon stems from the interplay between the structural transverse gliding asymmetry and the transverse nonreciprocal interaction between the scatterers that comprise the waveguide. In that sense this phenomenon may be regarded as the acoustic analog to one-way optical waveguiding that is based on the Voigt magneto-optical configuration such as in \cite{Mazor2015}.
%
%
%
%
%
Lastly, due to the similarity between the physical mechanisms, we expect that our proposed concept for one-way waveguides may be extended also to systems with transverse \emph{synthetic motion} along the waveguide cross section, (e.g., by space-time modulation), and thereby also to other physical realms such as electromagnetics and optics.

\begin{acknowledgments}
This research was supported by the Israel Science Foundation (grant No. 1353/19). 
Y.H would like to thank Prof. Eldad Avital for insightful discussions.  
\end{acknowledgments}


\begin{thebibliography}{29}
\bibitem{Comm1} 
To see that, we write the wave equation for the pressure field in a fluid with uniform mean flow velocity $U_0\hat{z}$,
\begin{equation}
(\partial_t + U_0\partial_z)^2 p = c^2\nabla^2 p.
\end{equation}
Considering propagation only on the $xz$ plane, its monochromatic plane wave solution is $p(x,z,t)=p_0e^{-j(k_xx+k_zz)}e^{j\omega t}$, and subject to the dispersion relation
\begin{equation}
k_z^2 + k_x^2 = k^2 - 2\frac{kU_0}{c}k_z + \frac{U_0^2}{c^2}k_z^2.
\end{equation}
Clearly, for a plane wave that propagates parallel to the $x$-axis [$z$-axis], $k_z=0$ [$k_x=0$], and therefore $k_x=\pm k$ [$k_z=k/(q_0\pm1)$], where $k=\omega/c$ is the wavenumver in the stationary medium and $q_0$ is the Mach number. Then, the phase velocity reads $v_p=\omega/k_{x,z}$, and thus $\tau_d^{A\leftrightarrow B}=L/v_p$.
\bibitem{Comm3}
Namely, the field that is back-reflected to the location of the 0'th element by the boundaries. See \cite{SM}.
\end{thebibliography}
\end{document}